\documentclass[preprint,12pt]{elsarticle}
\usepackage{amsmath, bm}

\usepackage{amssymb}

\usepackage{lineno}

\newcounter{bla}

\journal{Computer Physics Communications}

\begin{document}

\begin{frontmatter}



\title{Twister: Construction and structural relaxation of commensurate moir\'e superlattices}


\author[l1]{Saismit Naik}
\author[l2]{Mit H. Naik}
\author[l2]{Indrajit Maity}
\author[l2]{Manish Jain\corref{author}}

\cortext[author] {Corresponding author.\\\textit{E-mail address:} mjain@iisc.ac.in}
\address[l1]{Indian Institute of Science Education and Research, Pune 411008, India}
\address[l2]{Centre for Condensed Matter Theory, Department of Physics, Indian Institute of Science, Bangalore 560012, India}

\begin{abstract}
Introduction of a twist between layers of two-dimensional materials which leads to the formation of a 
moir\'e pattern is an emerging pathway to tune the electronic, vibrational and optical properties.
The fascinating properties of these systems is often linked to large-scale structural 
reconstruction of the moir\'e pattern. Hence, an essential first step in 
the theoretical study of these systems is the construction and structural relaxation of the 
atoms in the moir\'e superlattice.
We present the Twister package, a collection 
of tools that constructs commensurate superlattices for any combination of 2D materials and 
also helps perform structural relaxations of the moir\'e superlattice.
Twister constructs commensurate moir\'e superlattices using the coincidence lattice method
and provides an interface to perform structural relaxations using classical forcefields.
\end{abstract}


\end{frontmatter}



{\bf PROGRAM SUMMARY}

\begin{small}
\noindent
{\em Program Title:}    Twister                                \\
{\em Program obtainable from:}  https://github.com/qtm-iisc/Twister        \\
{\em Journal Reference:}                                      \\
{\em Catalogue identifier:}                                   \\
{\em Licensing provisions:} Open source BSD License                \\
{\em Programming language:}  Python                              \\
{\em Computer:} Any computer with Python3 installed. The package has been tested with Python3.6          \\
{\em Operating system:} Unix/Linux/Windows                    \\
{\em RAM:} 5-1000 MB (dependent on system size)           
{\em Classification:}                              \\
{\em External routines/libraries:} numpy, scipy, mpi4py, matplotlib \\
{\em Nature of problem:}        \\  
{\em Solution method:}          \\
{\em Running time:} 1-600 minutes (depends on the number of processors and system size)\\
\end{small}



\section{Introduction}

Two-dimensional (2D) materials, with their extensive spectrum of physical properties and 
long reaching applicability in technology, have triggered
a wave of innovation in the realm of material science. 
Twisting the layers of 2D materials to form moir\'e patterns is an 
emerging direction of research 
\cite{Nat.Cao2020,NPhys.Choi,NL.Haddadi,PRB.Chebrolu,Nano.xian,
NComm.Kennes,Nat.Regan,PRL.Lian,NatM.Wang,PRL.topo} 
which was triggered by the recent discovery of 
unconventional superconductivity and Mott-insulating phases in twisted
bilayer graphene \cite{Nat.Cao_1,Nat.Cao_2}. 
An important feature of the moir\'e patterns is that they undergo 
structural reconstructions as predicted by
theoretical simulations 
\cite{PRB.Carr_Soliton,2D.Oleg,2D.Wijk,arxiv.Maity,PRL.Naik,PRL.Enaldiev,arxiv.Xia,Nsc.Culchac,
PRB.Procolo,2D.Jain,arxiv.Leconte,PRB.Nam,arxiv.Lopez} 
and 
corroborated by experiments \cite{PNAS.Alden,ANano.Rosenberg,NMat.Yoo, NMat.Li,NNano.Weston,NatPhys.Woods,NNano.Mcgilly}. 
The atomic reconstructions can strongly modify the electronic structure
\cite{NMat.Yoo,arxiv.Maity,PRL.Kang,NNano.Weston,NMat.Li,PRR.Carr,PRL.Zhao,PRL.Naik,
PRB.Naik,NatPhys.Utama,arxiv.Angeli,arxiv.Xian} 
and 
vibrational modes \cite{PRB.Koshino,PRR.Maity,NSc.Debnath} of the moir\'e superlattice.


Most theoretical methods, particularly those based on classical forcefields or first-principles 
density functional theory \cite{PR.Kohn} (DFT), require the creation of a periodic simulation cell. Application of 
an arbitrary twist between the layers of a 2D material leads to an incommensurate moir\'e which is intractable 
for these methods. 
It is thus essential to find special twist angles \cite{PRB.Uchida,PRL.Lopes} which lead to moir\'e superlattices 
that are commensurate with periodic boundary conditions. 
A common approach to find these twist angles is the coincidence lattice method 
\cite{PRB.Uchida,JPCC.Koda}. In this method, 
the commensurate twist angle is determined as the one at which
perfect coincidence occurs for a set of lattice points in the top and bottom layers.
This method has been used to derive an analytical expression for the commensurate 
twist-angles in systems which have a hexagonal lattice and exactly same lattice constant in the top and bottom 
layer. To go beyond this restricted class to 2D materials, a general implementation of this method is necessary.
Furthermore, it is important to note that perfect coincidence of lattice sites is often not possible or 
 only possible at very large length scales. In this case a small strain may be applied 
 to each layer to obtain a commensurate superlattice of 
reasonable size \cite{JPCC.Koda}.

At large moir\'e length scales, all moir\'e superlattices undergo structural
reconstructions arising from in-plane and out-of-plane displacement of atoms from 
the rigidly-twisted structure. In-plane displacements lead to a change in the local
stackings giving rise to strain-soliton networks \cite{PRL.Naik,arxiv.Maity,PRB.Carr_Soliton,2D.Oleg,PNAS.Alden,NNano.Weston}
and out-of-plane displacements lead 
to a varying interlayer spacing and/or 
buckling of the layers \cite{NMat.Li,SA.Zhang,ANano.Waters,NL.Dai,Nat.Butz} in the moir\'e.
Hence, once the commensurate moir\'e superlattice is constructed, performing structural 
relaxation is the essential next step before exploring its electronic or optical properties.
Despite evidence of structural reconstructions, many theoretical studies of moir\'e superlattices 
do not take this into account \cite{PRL.topo,PRL.hubba,PRR.Pan,arxiv.Zhang,PRB.Bi,PRL.excit}.
This is because lattice relaxations using van der Waals corrected 
DFT is a major computational bottleneck due to the 
large number of atoms that constitute the moir\'e superlattice. 
An alternative 
computationally cheaper approach to perform the structural relaxation of the moire superlattices
is based on classical forcefields. This approach eliminates the electronic degrees of freedom and 
employs a theoretical description involving only nuclei. The interactions between nuclei 
are represented with simple inter-atomic potentials, whose parameters are obtained by 
fitting to accurate DFT calculations. 
For instance, the structural relaxations of the moir\'e superlattices of twisted bilayer of transition metal 
dichalcogenides (TMDs)
are performed with a combination of Stillinger-Weber \cite{PRB.SW} (SW) and registry-dependent 
Kolmogorov-Crespi \cite{PRB.KC} (KC) 
potentials. The SW and KC potential describes intralayer and interlayer interaction in these 
systems. 
The SW potential parameters were obtained by fitting to DFT-derived phonon dispersion 
of single layer TMD \cite{Chap.SW}, whereas the KC potential parameters were obtained by 
fitting interlayer 
binding energy landscape to van der Waals corrected DFT results \cite{JPCC.Naik, JCTC.Leven, JPCC.Maaravi, DRIP,JCP.Leven}. 
The fit forcefields can be used to obtain moir\'e reconstructions in good agreement with DFT relaxations \cite{JPCC.Naik}
at a fraction of the computational cost. 
However, the technical details of classical force-field based simulation are
non-trivial. An all-inclusive package that constructs and automates the performance of 
lattice relaxations of the moir\'e superlattice can bridge this gap for non-experts.



We present the Twister package to construct commensurate moir\'e
superlattices for any 2D homobilayer or heterobilayer material. 
Twister helps find minimum-area commensurate superlattices using 
the coincidence lattice method for a given range of twist angles 
and user-defined strain tolerance. 
The superlattice vectors and atoms within the superlattice are 
written to a file for further study of its structural and electronic properties. The construction of
commensurate superlattice has been tested for various homobilayer and heterobilayer cases: graphene/hBN, 
MoS$_2$/MoSe$_2$, hBN/MoSe$_2$, twisted bilayer phosphorene and MoS$_2$. 
We use the package to find a range of commensurate twist angles for the case 
of MoS$_2$/MoSe$_2$ and twisted bilayer phosphorene. Twister further also 
streamlines the process of studying moir\'e reconstruction using classical forcefield calculations as 
implemented in the LAMMPS \cite{JCP.Plimpton} package. 
The streamlined process is demonstrated for 
twisted bilayer MoS$_2$, twisted bilayer graphene and twisted MoS$_2$/MoSe$_2$ moir\'e patterns.

\section{Construction of commensurate superlattices}

The method used to construct the commensurate superlattices depends on the 
type of lattice vectors in the bottom and top layer forming the moir\'e. If the 
two lattices are hexagonal and with identical lattice constants, we use an 
analytical expression \cite{PRL.Lopes} to obtain the commensurate twist-angles ($\theta_i$).
\begin{equation} \label{eqn_i}
	cos(\theta_i) = \frac{3i^2 + 3i + 0.5}{3i^2 + 3i + 1}, 
\end{equation}
where $i$ is a whole number. The corresponding superlattice vectors are given by: 
$\mathbf{A}_1 = i \mathbf{a}_1 + (i+1) \mathbf{a}_2$, $\mathbf{A}_2 = -(i+1) \mathbf{a}_1 + (2i+1) \mathbf{a}_2$
where $\mathbf{a}_1$ and $\mathbf{a}_2$ are the unit-cell lattice vectors. 
In the superlattices generated using this equation, the peridocity of the stackings 
matches the moir\'e unit-cell periodicity. Ie. each type of stacking (say AA) is 
present once per moir\'e unit-cell.
For commensurate twist-angles in addition to $\theta_i$, a general form for 
commensurate twist-angles is given by \cite{PRB.Uchida}:
\begin{equation}  \label{eqn_mn}                                                                                    
	cos(\theta_{m,n}) = \frac{n^2 + 4nm + m^2}{2(n^2 + mn + m^2)},
\end{equation}
where m, n are whole numbers. The superlattice vectors are: $\mathbf{A}_1 = m \mathbf{a}_1 + n \mathbf{a}_2,
\mathbf{A}_2 = -n \mathbf{a}_1 + (m+n) \mathbf{a}_2$. This general equation yields all possible commensurate 
superlattices and not just ones with minimum area. The superlattice can have mutiple AA stackings per moir\'e 
unit-cell. 

\begin{figure}[h]
 \centering
  \includegraphics[scale=0.5]{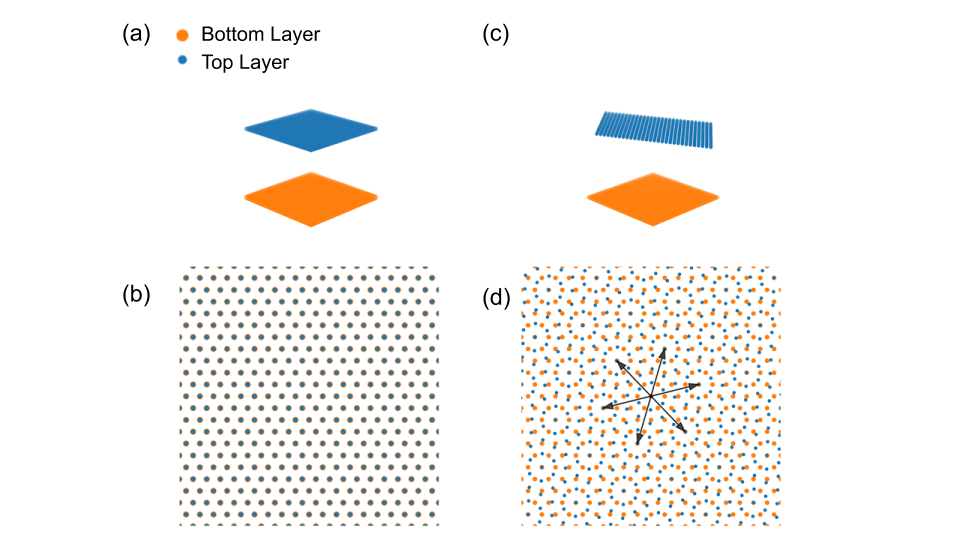}
  \caption{ a) and c) 3-D representations of untwisted and twisted  MoS$_2$ homobilayer respectively. 
	Twist angle is 27.8$^\circ$ in c). 
	b) and d) Top view of a) and c), respectively.
	Six solution vectors of Eqn. \ref{eqn_cslt} (with minimum superlattice area) are shown in d)}
\label{figure:1}  
\end{figure}

For a combination of layers with dissimilar lattices we use the coincidence site lattice theory to 
generate the superlattice. Consider the unit-cell lattice vectors of bottom layer 
to be $\mathbf{a}_1$, $\mathbf{a}_2$, $\mathbf{a}_3$ 
and of top layer to be $\mathbf{b}_1, \mathbf{b}_2, \mathbf{b}_3$. 
Let $\mathbf{R}_\theta$ be the rotation matrix corresponding to a rotation angle $\theta$ in the xy plane. 
Top layer is rotated about the origin keeping the bottom layer fixed. The rotated lattice vectors, 
$\mathbf{b}^r_i$ are given by:

\begin{equation}
	\mathbf{b}_i^r = \mathbf{b_i}  \mathbf{R}^\mathbf{T}_{\theta}
\end{equation}
Coincidence of a subset of lattice sites occurs when a lattice vector in bottom layer, 
$\mathbf{v}_1 = n_1\mathbf{a}_1 + n_2\mathbf{a}_2$, 
superposes with a lattice vector in the rotated top layer, 
$\mathbf{v}_2^r = m_1\mathbf{b}^r_1 + m_2\mathbf{b}^r_2$, ie.

\begin{equation} \label{eqn_cslt}
	n_1\mathbf{a}_1+n_2\mathbf{a}_2 = m_1\mathbf{b}_1^r+m_2\mathbf{b}_2^r
\end{equation}
Figure \ref{figure:1} graphically represents the above steps for 27.8$^\circ$ twisted bilayer MoS$_2$. 
The shortest coinciding vectors have been plotted and any two non-parallel 
vectors can be equivalently used as superlattice vectors for the given twist angle.

For heterobilayer materials, it is often difficult to obtain exact 
coincidence since lattice parameters of the constituent layers are different. Hence, we introduce 
a user-defined upper limit of mismatch, $\Delta$, between the vectors in $\mathrm{\AA}$ units, 
\begin{equation} \label{eqn_clt_delta}
	|\mathbf{v}_1 - \mathbf{v}_2^r| \leq \Delta
\end{equation}

It is important to find superlattices with as small a mismatch as possible to ensure that the 
superlattice is commensurate with the twist-angle. A large value of $\Delta$ can introduce spurious 
strains in the superlattice. For an arbitrary combination of layers, the value of mismatch 
or the size of superlattice is often too large despite 
sweeping a large range of twist-angles. In these cases, more effort is needed to find perfectly 
commensurate superlattices. In the following subsections, we describe two methods to improve coincidence.

After the computation of all vectors co-incident within the mismatch, pairs of solutions
($\{n_1\mathbf{a}_1+n_2\mathbf{a}_2, n_1'\mathbf{a}_1+n_2'\mathbf{a}_2\}$) are checked to be 
non-parallel by ensuring their vector cross products is non-zero. 
The solutions can also be further restricted to have a specific angle between the vectors.
In the final step, we choose superlattices with minimum area with a variation of 1\% allowed around 
the minima. The final output is a list of appropriate superlattice vectors and 
the corresponding strained lattice parameters. 

\subsection{Straining lattice parameters}

\begin{figure}[h]
 
  \includegraphics[scale=0.4]{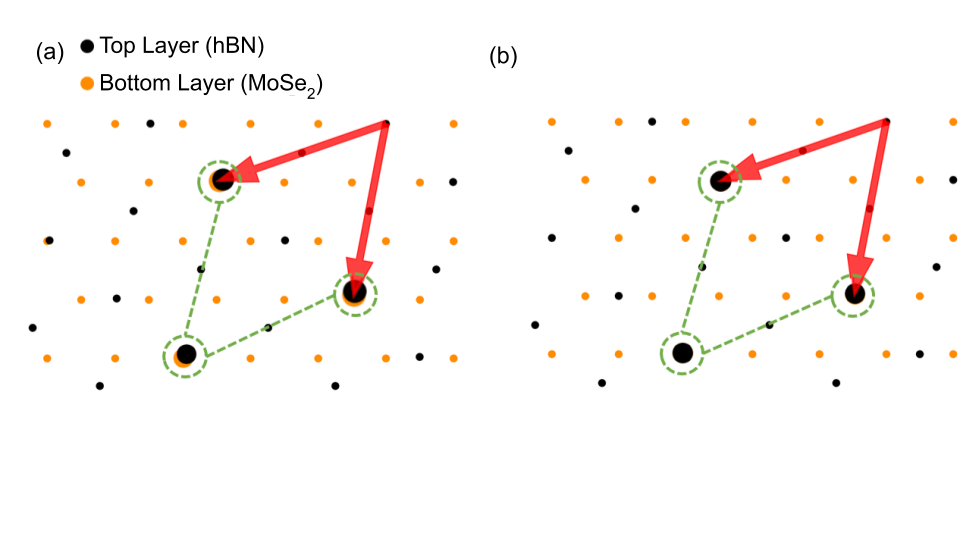}
  \caption{ Red arrows are the superlattice vectors of hBN/ MoSe$_2$ bilayer twisted at 
	19.1$^\circ$. The corner lattice points are encircled. a) Supercell with unstrained lattice 
	parameters. b) Supercell with strained parameters with improved overlap.
  }
\label{figure:2}  
\end{figure}

The lattice parameters of each layer can be uniformly strained as
demonstrated for the case of hBN/MoSe$_2$ in Fig. \ref{figure:2}.
A biaxial strain ($\beta$) is applied to each layer such 
that the following equation is satisfied \cite{JPCC.Koda}:

\begin{equation}
	(1+\beta)|\mathbf{v}_1| \approx (1-\beta)|\mathbf{v}_2^r|
\end{equation}

Lattice parameters of one layer are 
strained by $\beta$ and those of the lower layer by $-\beta$ to improve coincidence.
In certain situations, it is desirable to allow for only one layer to be strained. If only lattice parameters 
of the top layer are to be strained, $\beta$ is applied to the top layer such that:

\begin{equation}
|\mathbf{v}_1| \approx (1+\beta)|\mathbf{v}_2^r|
\end{equation}

If only lattice parameters of the bottom layer are to be strained, $\beta$ is applied to the bottom layer such 
that:

\begin{equation}
(1+\beta) |\mathbf{v}_1| \approx |\mathbf{v}_2^r|
\end{equation}

Two values of $\beta$ are generated for each pair of solutions ($\{n_1\mathbf{a}_1+n_2\mathbf{a}_2, n_1'\mathbf{a}_1+n_2'\mathbf{a}_2\}$)
 and the averaged value is used to form the superlattice vectors. The mismatch after straining the superlattice vectors is 
 computed as demonstrated in Eqn. \ref{eqn_clt_delta}.

\subsection{Applying a strain tensor}

When straining the lattice parameters alone does not yield small mismatch, the  
unit-vectors of the top layer can be deformed to ensure perfect coincidence.
However, this may change the angle between the lattice vectors of the strained layer leading to 
breaking of symmetries within that layer. The deformation tensor ($\bm{D}$) is computed such
that:

\begin{equation}
	\begin{bmatrix} n_1 & n_2 \\ n_1' & n_2'\end{bmatrix} \begin{bmatrix} 
    \mathbf{a}_1 \\ \mathbf{a}_2 \end{bmatrix} = 
    \begin{bmatrix} m_1 & m_2 \\ m_1' & m_2'\end{bmatrix} \begin{bmatrix} \mathbf{b}_1 \\ \mathbf{b}_2 \end{bmatrix} \bm{D}  \bm{R^T_{\theta}} 
\end{equation}

Any deformation of a lattice can be expressed in terms of a pure rotation of the lattice and a 
symmetric strain tensor. Through a polar decomposition \cite{Horn}, 
we have $\bm{D} = \bm{\epsilon} \bm{R_\phi}$:

\begin{equation} \label{eqn_def}
\begin{bmatrix} n_1 & n_2 \\ n_1' & n_2'\end{bmatrix} \begin{bmatrix} 
\mathbf{a}_1 \\ \mathbf{a}_2 \end{bmatrix} = 
\begin{bmatrix} m_1 & m_2 \\ m_1' & m_2'\end{bmatrix} \begin{bmatrix} 
\mathbf{b}_1 \\ \mathbf{b}_2 \end{bmatrix} \bm{\epsilon} \bm{R_\phi}  \bm{R^T_{\theta}}
\end{equation}

The pure rotation leads to a further change in the twist-angle by $\phi$. The change in the twist-angle along 
with a strain in the unit-vectors (by $\bm{\epsilon}$) leads to a perfectly commensurate superlattice.

\section{Test Systems}

The construction of commensurate superlattices using our implementation 
(Eqn \ref{eqn_cslt}) is verified for five test systems --
twisted bilayer MoS$_2$, MoS$_2$/MoSe$_2$ heterobilayer, 
Graphene/hBN heterobilayer and hBN/MoSe$_2$ heterobilayer 
to compare with Ref. \cite{JPCC.Koda}.
To find the commensurate superlattices (using the procedure described in section 2) 
we use the following constraints.
For MoS$_2$/MoSe$_2$, MoS$_2$ $/$ MoS$_2$ and hBN$/$MoSe$_2$, the angle between superlattice 
vectors is fixed at 60$^\circ$. For Graphene$/$hBN, the angle between superlattice vectors is fixed to 
be 120$^\circ$. Maximum strain, $\beta$, allowed in the lattice parameters is 1\%. 
Mismatch ($\mathrm{\AA}$) threshold, $\Delta$, is set to be less that 0.1 $\mathrm{\AA}$.
The commensurate superlattices, number of atoms and 
strains are in agreement with the reported results \cite{JPCC.Koda,PRL.Kang}.
The lattice parameters used in our simulation and the
results are shown in Table \ref{table:1} and Table \ref{table:2}, respectively.
The input files along with instructions on how to use the code to generate the
commensurate superlattices for these test systems is included in the Twister package.

\begin{table}[h!]
\centering

\begin{tabular}{c@{\hskip 0.14in}r@{\hskip 0.08in}r}
\hline
\hline
	Materials & Lattice parameter ($\mathrm{\AA}$) \\
\hline

MoS$_2$  & 3.164\\

MoSe$_2$ & 3.301\\

Graphene & 2.467 \\

hBN      & 2.512 \\

Phosphorene  & 3.321, 4.62   \\

\hline

\hline

\end{tabular}

\caption{
Lattice parameters of each material
}
\label{table:1}
\end{table}

\begin{table}
\centering
\resizebox{\columnwidth}{!}{%
\begin{tabular}{c@{\hskip 0.14in}r@{\hskip 0.08in}r@{\hskip 0.08in}r@{\hskip 0.08in}r@{\hskip 0.08in}r@{\hskip 0.08in}r}
\hline
\hline
Materials & Twist angle & Superlattice vectors &   Atoms &   Biaxial strain \\
	  &             & ($n_1$, $n_2$), ($n'_1$, $n'_2$)    &            &             \\
\hline

	MoS$_2$$/$MoS$_2$ & 27.8$^\circ$ & (-4, 3), (-3, -1)  & 78 & $\pm$0.0\\

	MoS$_2$$/$MoSe$_2$ & 16.1$^\circ$ & (-1, -3), (3, -4)  & 75 & $\pm$0.119\\

	Graphene$/$hBN   & 0.0$^\circ$ & (-37, 0), (0, 37)  & 5330 & $\pm$0.466\\

	hBN$/$MoSe$_2$   & 19.1$^\circ$  & (1, -3), (3, -2)   & 26  & $\pm$0.333\\

\hline

\hline

\end{tabular}
}
\caption{ Results obtained from Twister for the test systems to compare with the 
          commensurate twist-angles reported in Ref. \cite{JPCC.Koda}.
	}
\label{table:2}
\end{table}

\section{Examples}

\subsection{Commensurate twist-angles}

Apart from the above test examples, we use Twister to compute
commensurate twist-angles for the
heterobilayer systems of MoS$_2$/MoSe$_2$ and bilayer phosphorene. 
For systems of MoS$_2$/MoSe$_2$, the angle between superlattice vectors 
is fixed at 60$^\circ$. Strain for lattice parameters is kept below 1\% 
and distributed over both layers. Mismatch threshold is
$10^{-5} \mathrm{\AA}$. The size of the moir\'e superlattice is expected to 
increase as a function of reducing twist-angle. 
We accordingly increase the search range of $n_1, n_2, m_1, m_2$ (Eqn. \ref{eqn_cslt}) as a function of decreasing 
twist-angle. 
The commensurate twist-angles for MoS$_2$/MoSe$_2$ can be 
found in Table \ref{table:3}.
The commensurate twist-angles for twisted bilayer phosphorene are generated using the deformation 
tensor method described above. One of the layers is allowed to deform to form a perfectly 
commensurate superlattice. A fine tuning of the twist-angles is 
not necessary for this method. For example, starting with 
$\theta = 6.9^\circ$ leads to the following strain and rotation
matrices (Eqn. \ref{eqn_def}):
\begin{equation}
\bm{\epsilon} = \left[ \begin{array}{cc} 0.992893  & -0.001897  \\ -0.001897  & 0.993364 \end{array}\right];
\bm{R_\phi}  = \left[ \begin{array}{cc} 0.999995 &  -0.003102 \\ 0.003102 & 0.999995 \end{array}\right]
\end{equation}

The $\bm{\epsilon}$ tensor is applied to the lattice vectors and 
the $\bm{R_\phi}$ leads to a change in the twist-angle between the layers 
by $\phi = -0.17773^\circ$. The commensurate twist-angle is thus, 
$\theta + \phi = 6.72227^\circ$. To test the implementation, we  
also start with $\theta = 6.5^\circ$ and find a solution:
\begin{equation}
\bm{\epsilon} = \left[ \begin{array}{cc} 0.992919  & -0.001900  \\ -0.001900  & 0.993338 \end{array}\right];
\bm{R_\phi}  = \left[ \begin{array}{cc}  0.999992 &  0.003879 \\ -0.003879 & 0.999992 \end{array}\right]
\end{equation}

The strain tensor is nearly the same, 
while the change in twist-angle, $\phi = 0.2222699^\circ$. 
As expected, the final commensurate twist-angle ($\theta + \phi$) 
remains identical for 
different starting values of $\theta$. 
The commensurate twist-angles for twisted bilayer phosphorene generated 
using this method are provided in Table \ref{table:4}.
The input files, along with instructions on how to run 
the code, are provided with the Twister package.
More details on using the code and the workflow is 
described in the next section.

\begin{table}[h!]
\centering
\begin{tabular}{c@{\hskip 0.14in}r@{\hskip 0.08in}r@{\hskip 0.08in}r@{\hskip 0.08in}r@{\hskip 0.08in}r@{\hskip 0.08in}r}
\hline
\hline
Twist Angle & Atoms & Superlattice vectors & Strain (\%) & Mismatch ($\mathrm{\AA}$)  \\
            &       & (n1, n2), (n1', n2') &  &   &           \\
\hline
0.0$^\circ$ & 3315 & (24, 0), (0, 24) &\ 0.00856 &\  $10^{-9}$ \\

1.56654$^\circ$ & 2283 & (11, -23), (-12, -11)&\ 0.0501 &\  $5\times10^{-6}$   \\

2.84558$^\circ$ & 1362 & (4, -17), (17, -13)&\ 0.0847&\  $2\times10^{-6}$\\

3.10565$^\circ$ & 1248 & (-16, 13), (-13, -3)&\ 0.0454&\ $ 2\times10^{-6}$ \\

4.715$^\circ$ & 696 & (0, -11), (11, -11)&\ 0.0371 &\  $2\times10^{-6}$\\

7.3111$^\circ$  & 327 & (-8, 1), (-7, 8)&\ 0.176&\ $ 4\times10^{-6}$ \\ 

11.17856$^\circ$  & 456  &(3, -10), (10, -7) &\ 0.145&\ $1\times10^{-6}$\\

14.11315$^\circ$  & 384  & (7, -9), (-2, -7)&\ 0.226&\ $ 2\times10^{-6}$ \\

16.1021$^\circ$  & 75  & (-4, 1), (-3, 4)&\ 0.118 &\ $3\times10^{-6}$ \\

21.45415$^\circ$  & 384  & (7, -9), (-2, -7)&\ 0.226 &\ $10^{-6}$ \\

\hline

\hline

\end{tabular}

\caption{Commensurate twist-angles in the range 0-30$^\circ$ generated for MoS$_2$/MoSe$_2$ 
	using Twister. Lattice parameters were strained to improve mismatch.
}
\label{table:3}
\end{table}

\hspace{10 mm}

\begin{table}[h!]
\centering

\begin{tabular}{c@{\hskip 0.14in}r@{\hskip 0.08in}r@{\hskip 0.08in}r@{\hskip 0.08in}c@{\hskip 0.08in}c@{\hskip 0.08in}r}
\hline  
\hline
Twist Angle & Atoms & Superlattice vectors   & Strain tensor    &   Mismatch ($\mathrm{\AA})$       \\
            &      & (n1, n2), (n1', n2')&     &      &          \\
\hline
&\  &\ &\ &\ \\
1.595323$^\circ$ & 10444 &(45, 0), (0, 29) &\  $\left[ \begin{array}{cc} 1.00047 & 0.00306  \\ 0.00306 & 1.00030 \end{array} \right]$  &\ $0.0$  \\
&\  &\ &\ &\ \\
2.760976$^\circ$ & 3460 & (27, 0), (0, 16)&\ $\left[ \begin{array}{cc} 0.99901 & 0.00329 \\ 0.00328 & 0.99869 \end{array} \right]$ &\ $0.0$  \\
&\  &\ &\ &\ \\
3.523156$^\circ$ & 2116 & (22, 0), (0, 12)&\  $\left[ \begin{array}{cc} 0.99821 & 0.00166 \\ 0.00166 & 0.99801 \end{array} \right]$ &\ $0.0$  \\ 
&\  &\ &\ &\ \\
4.265478$^\circ$ & 1444 & (18, 0), (0, 10)&\ 	$\left[ \begin{array}{cc} 0.99744 & 0.00268 \\ 0.00268 & 0.99704 \end{array}\right]$ &\ $0.0$  \\
&\  &\ &\ &\ \\
5.404710$^\circ$  & 900 & (14, 0), (0, 8)&\ 	$\left[ \begin{array}{cc} 0.99602 & 0.00469 \\ 0.00469 & 0.99513 \end{array}\right]$ &\ $0.0$  \\
&\  &\ &\ &\ \\
6.722269$^\circ$  & 580  & (12, 0), (0, 6)&\ 	$\left[ \begin{array}{cc} 0.99289 & -0.00189 \\ -0.00189  & 0.99337 \end{array}\right]$ &\ $0.0$  \\
&\  &\ &\ &\ \\
\hline

\hline

\end{tabular}

\caption{Commensurate twist-angles of bilayer black phosphorous generated using Twister. 
A strain tensor is applied to one of the layers to improve mismatch. }

\label{table:4}
\end{table}

\hspace{10 mm}

\subsection{Structural relaxations}
\begin{figure}[h]
	 \centering                                                                                                     
	   \includegraphics[scale=0.24]{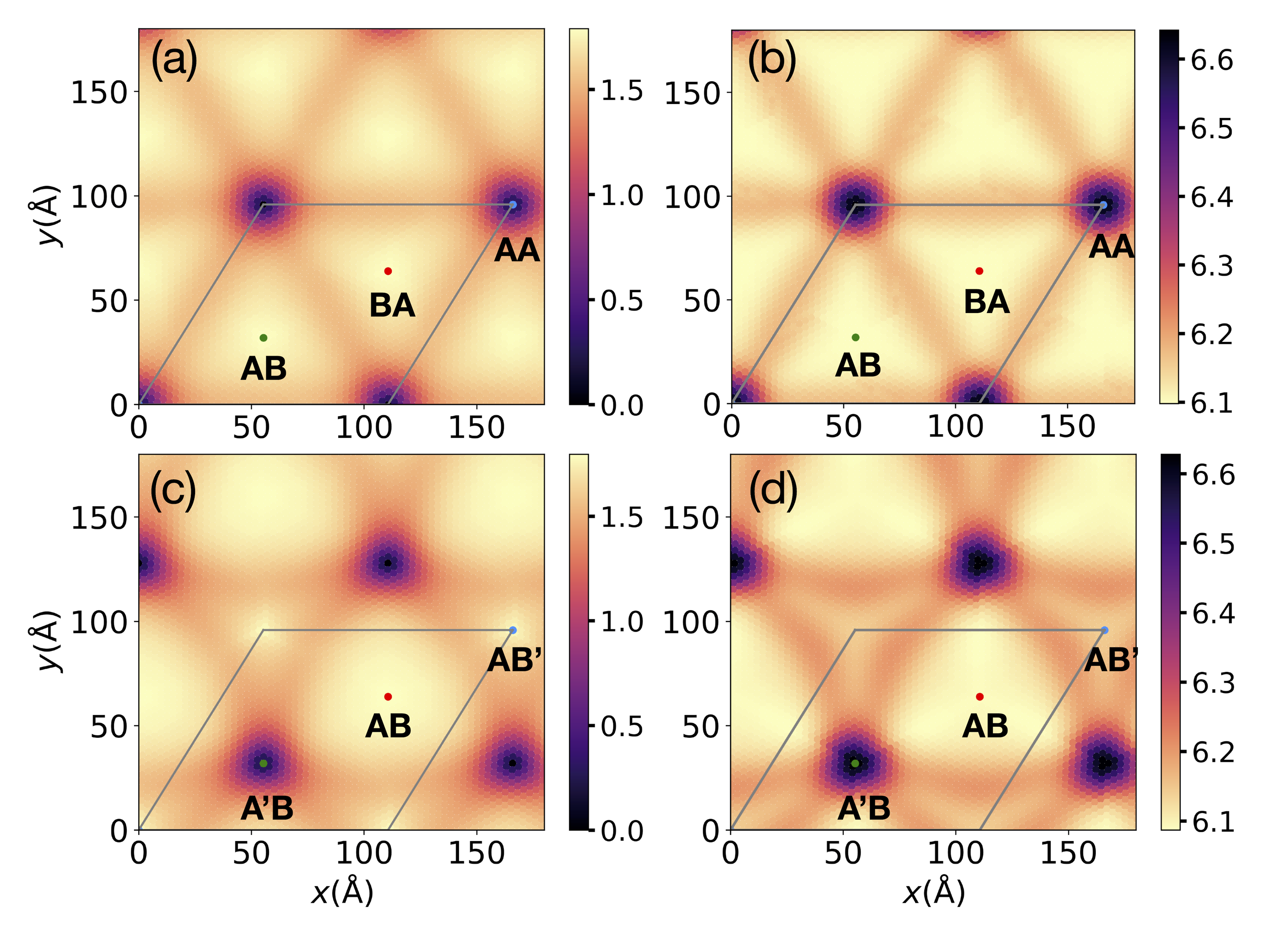}  
	     \caption{ a and b (c and d) Order-parameter distribution and interlayer spacing distribution in 
	     1.61$^\circ$ (58.39$^\circ$) twisted bilayer MoS$_2$, respectively. 
	       }                                                                                                             
	       \label{opils}  
\end{figure}

\begin{figure}[h]
	 \centering                                                                                                     
	   \includegraphics[scale=0.32]{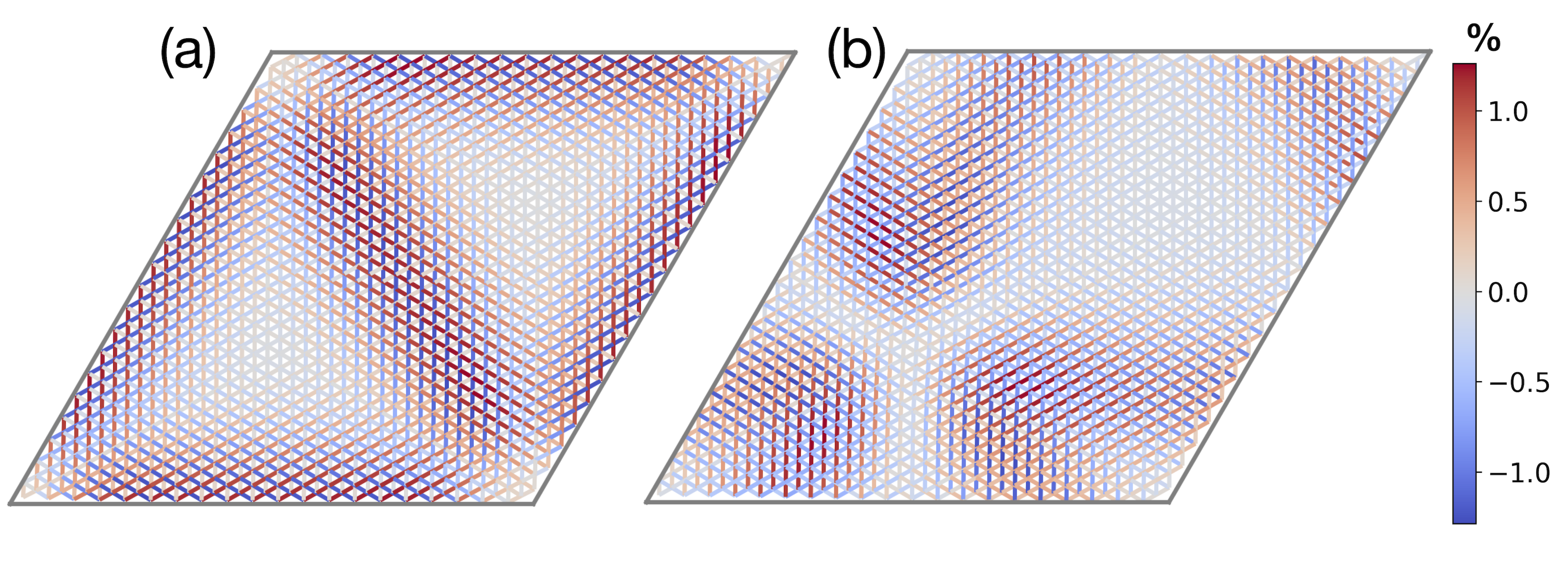}
	     \caption{ a and b Distribution of strain in the bottom layer of 1.61$^\circ$ and 58.39$^\circ$) twisted bilayer 
	     MoS$_2$. Lines are drawn between every Mo atom and its 6 nearest neighbors. The color of the line represents 
	     the strain in that direction.
	       }                                                                                                             
	       \label{strains}                                                                                                
\end{figure}

We demonstrate the automation of
construction and relaxation of moir\'e 
superlattices (using the LAMMPS package) for the case of twisted bilayer MoS$_2$. 
Using a single script, \emph{homo.sh}, a range of commensurate superlattices
can be generated and relaxed. Using this
script we perform the following: 
\begin{enumerate}
	\item A file, \emph{hex.table}, which contains a series of twist
angles and superlattice vectors is created. 
\item  The moir\'{e} superlattice and atom positions are generated for each twist-angle in this file. 
\item To perform the relaxation using LAMMPS, the superlattice vectors must be 
reoriented so that one of the superlattice vectors is along the $(100)$ direction.
Consider three superlattice vectors of the moir\'{e} superlattice generated using
Eqn. \ref{eqn_i}: $\mathbf{v}_1, \mathbf{v}_2$ and $\mathbf{v}_3$. 
	$\hat{\mathbf{v}}_3$ is along the out-of-plane, $z$ direction. 
The transformation is thus computed in the following manner, 
$$\vec{A}_1=\lVert \vec{v}_1 \rVert \ \hat{x}$$ 
$$\vec{A}_2=(\vec{v}_2\cdot \hat{v}_1)\ \hat{x} + (|\hat{v}_1 \times \vec{v}_2|) \ \hat{y}$$                                      
$$ \vec{A}_3 = \vec{v}_3 $$ 
\item With the user-provided force-fields, the atomic relaxations are performed using the LAMMPS package for all the 
twist-angles in \emph{hex.table}. 
\end{enumerate}
The range of twist-angles in \emph{hex.table}
can be generated using both Eqn. \ref{eqn_i} or \ref{eqn_mn} according to the user's preference. 
The relaxed structures can be visualized with the OVITO \cite{ovito} software directly.
The order-parameter and interlayer spacing distribution for relaxed 1.61$^\circ$ and 58.39$^\circ$ 
twisted bilayer MoS$_2$ is shown in 
Fig. \ref{opils}. The order-parameter is defined as the shortest displacement of the top layer that transforms 
a given local stacking to the highest energy stacking in the moir\'e -- AA for 1.61$^\circ$ twist and 
A'B for 58.39$^\circ$ twist. The relaxation leads to increase in the area of low-energy stackings 
(AB, BA, AA', AB') and 
formation of shear-strain solitons at the boundary between adjacent low-energy stackings 
\cite{PRR.Maity,PRL.Naik, PRB.Naik,PRB.Carr_Soliton}.
The strain distribution in the bottom layer of the relaxed 
moir\'e pattern is shown in Fig. \ref{strains}. The strains are localised along the solitonic networks.

\section{Workflow}
In this section the sequential steps for operating the code are described. The code can be divided into three sections.

\subsection{Finding commensurate angle}
This part of the package is used to generate the commensurate twist-angles and the 
corresponding lattice vectors using the method described in section 1. The 
code, \emph{get$\_$ang.py}, is used for this purpose and uses an input file: \emph{get$\_$ang.inp}.

\begin{enumerate}
	\item Prepare input file, \emph{get$\_$ang.inp}. The lattice vectors for the top and 
bottom layer are provided by the user in this file. The basis atom positions are not necessary at this 
stage.
The input file is also 
designed to contain an exhaustive range of constraints for 
the superlattice to be found. The user can fix the angle between the superlattice vectors, 
the type and amount of strain applied on the layer parameters, the tolerance threshold, $\Delta$. The user 
also provides a range of twist-angles to be searched through this input file. On running
the code, a plot is generated with the lattice points and the superlattice vectors.
Vectors of the same color show super-cells of the same 
area. An output file is generated for every twist-angle in the range provided by the user and
contains all the solutions found by solving Eqn. \ref{eqn_clt_delta}.

	\item If too many solutions are found, the parameters can be changed to reduce the number of vectors and plot them for
confirmation. Usually, it involves reducing ranges of $n_1, n_2, m_1, m_2$ to look for smaller superlattices or
adjusting the strain
percent and mismatch allowed. 
If no superlattice vectors are found, changes to the input are suggested.

	\item Pick the appropriate super-cell vectors from the output and prepare the second input file, 
\emph{twist.inp}, to generate the atom positions.
\end{enumerate}

\subsection{Atoms in superlattice}
This part of the package, \emph{twister.py},
generates the atom coordinates in the moir\'e superlattice. 
The input files needed are $twist.inp$ and basis atom positions of the unit-cell for top and 
bottom layer. The code also verifies that the atoms lying on the boundary of the supercell 
are not missed or counted twice. The code verifies
that the number of atoms computed matches the expected value from area ratios. The expected 
number of atoms in top layer is given by: $n_{top} = n_b*\mathrm{Ar}_{sl}/\mathrm{Ar}_{uc}$, where $n_b$ is the 
number of basis atoms in top layer, $\mathrm{Ar}_{sl}$ is the area of the superlattice and $\mathrm{Ar}_{uc}$ is the 
area of the unit-cell. The output file, \emph{superlattice.dat}, contains 
the superlattice vectors and the atom coordinates.

\begin{figure}[h]
 \centering
  \includegraphics[width=\textwidth]{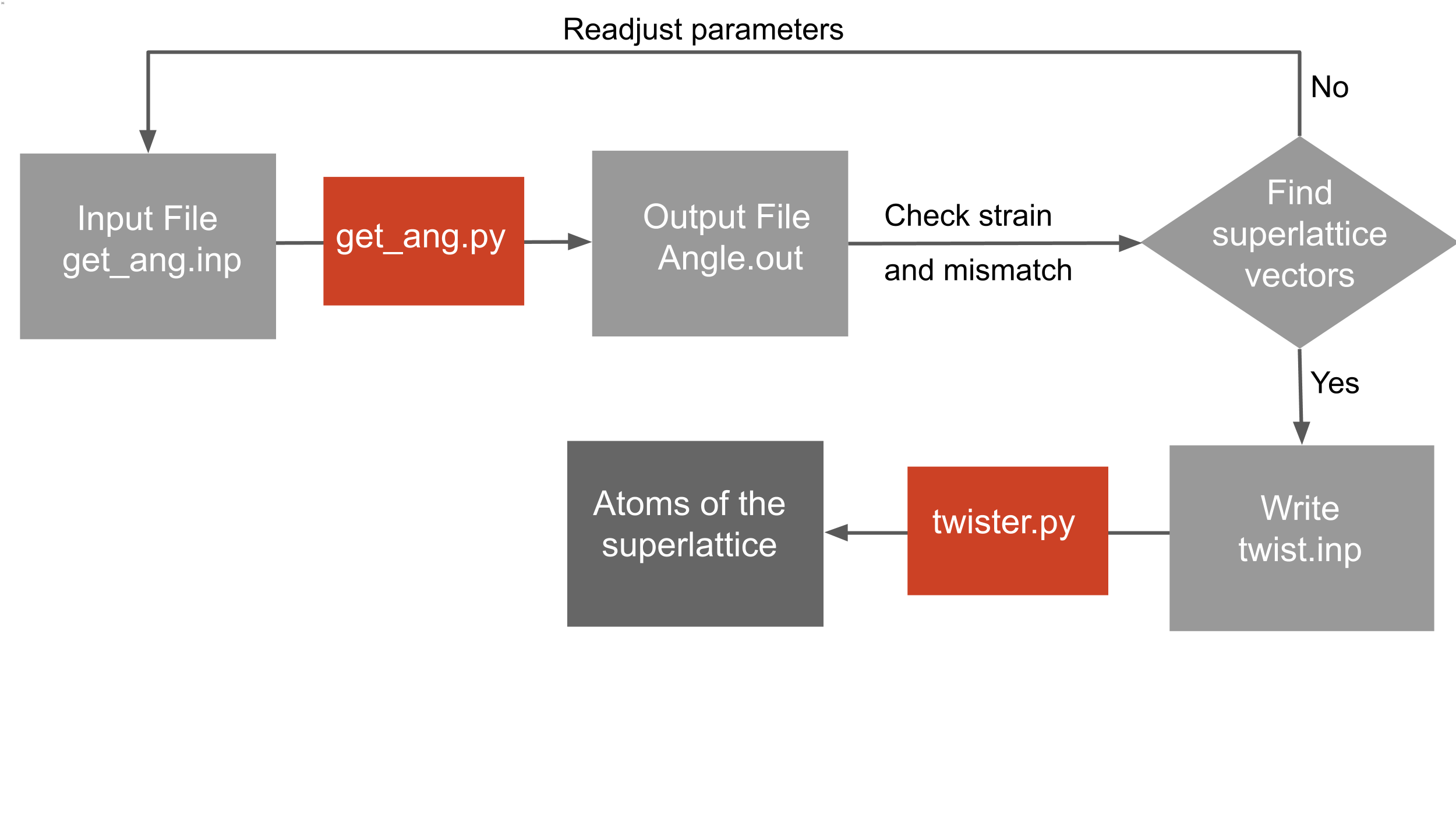}
  \caption{ The program flow for generating the supercell lattice and number of atoms for given material and twist angle.
  }
\label{figure:4}  
\end{figure}

A flowchart summary of all the above steps can be found in Figure \ref{figure:4}.

\begin{figure}[h]
 \centering
  \includegraphics[scale=0.46]{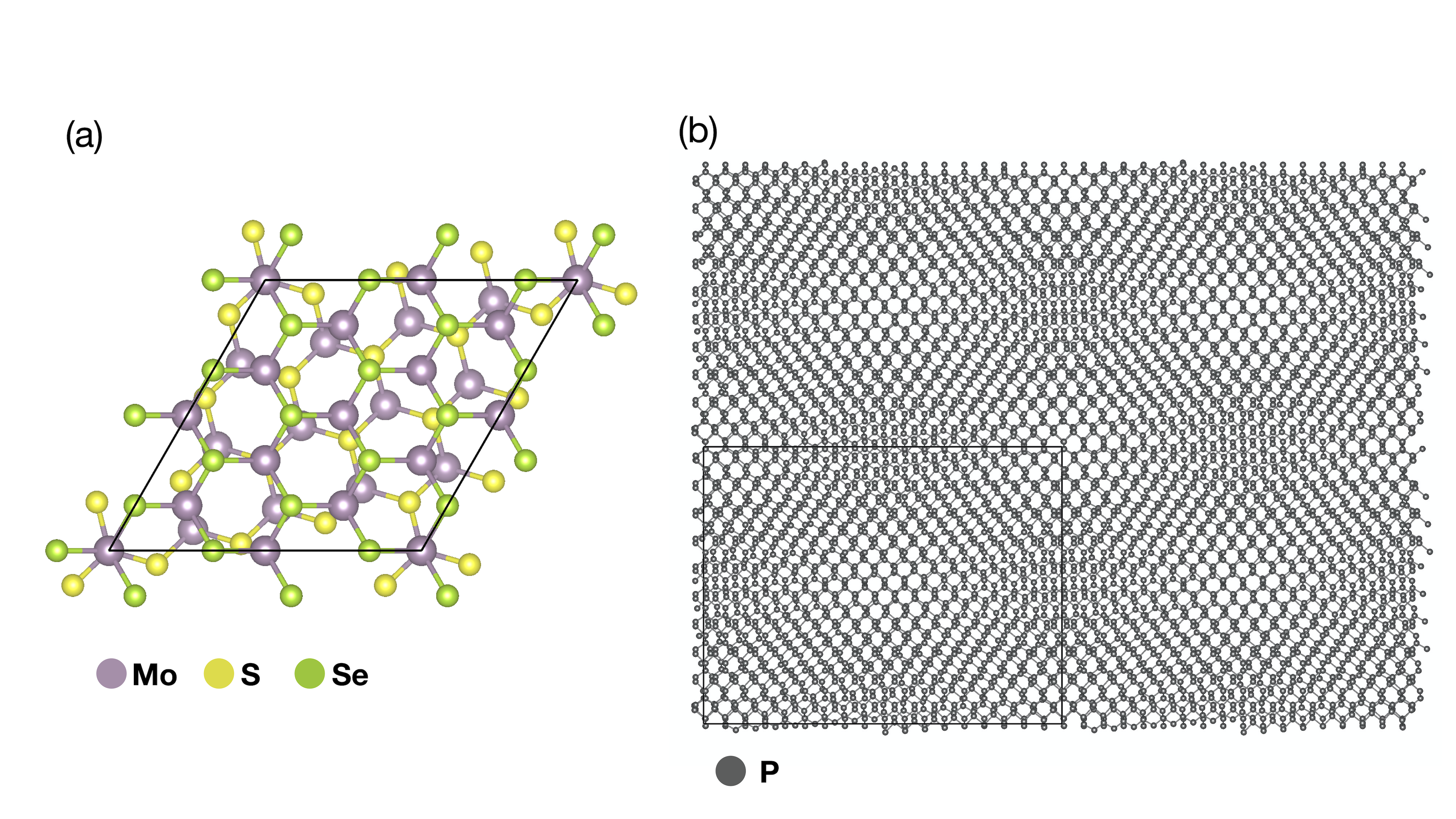}
  \caption{ 
  Superlattices of 
  a) 16.1$^\circ$ twisted MoS$_2$/MoSe$_2$ and
  b) 4.265$^\circ$ twisted bilayer
  black phosphorous.
  }
\label{figure:5}
\end{figure}
 
\section{Twister code framework}

\subsection{Input file}
We provide an example input file, \emph{get\_ang.inp} for the 
code \emph{get\_ang.py} described in the previous section to showcase the 
various options available to the user.

\begin{verbatim}
range_nm:
-20 21
celldm_a:
3.164 3.164 25.0
a1: 
0.5 0.8660254 0.0 
a2:
-0.5 0.8660254 0.0
a3:
0.0 0.0 1.0
Number_basis_atoms_a:
3
celldm_b:
3.301 3.301 25.0
b1:
0.5 0.8660254 0.0                                                                                            
b2:
-0.5 0.8660254 0.0                                                                                            
b3:                                                                                                           
0.0 0.0 1.0                                                                                                   
Number_basis_atoms_b:                                                                                      
3                                                                                                             
theta_range:                                                                                                 
16.05 16.15 0.01                                                                                              
mismatch (Angstrom):                                                                                          
0.0271                                                                                                          
strain_per:                                                                                                  
1.0         
strain_tensor_vector:
'False'
strain_layer:                                                                                                
'Both'                                                                                                   
fix_ang:                                                                                                     
'True'                                                                                                        
f_ang:                                                                                                       
60                                                                                                            
plot:
'Y'
\end{verbatim}

The unit of length is Angstrom ($\mathrm{\AA}$). Lattice parameters, unit-cell vectors 
and basis atoms for the top layer are provided in the line following the key-word: celldm\_a,(a1,a2,a3) and 
Number\_basis\_atoms\_a respectively and similarly for the bottom layer.
Lower bound, upper bound and step size for the twist-angle range is given by the string below 
theta\_range in the same order. Threshold for mismatch between coincident vectors of top and 
bottom layer can be specified in the string below mismatch (Angstroms).  
The user can apply a strain tensor by specifying 'True' below 
strain\_tensor\_vector otherwise, if specified 'False', 
only the lattice parameters (of the layer specified under strain\_layer) will be strained. 
Threshold for percentage 
strain allowed to improve mismatch is given below strain\_per. 
When entry under strain\_tensor\_vector is 'True', the deformation tensor 
method described above will be used.
Fixing angle between superlattice vectors and specifying the angle between superlattice vectors
are allowed by fix\_ang and f\_ang respectively. The user can plot the resultant superlattice 
vectors for each angle by 
specifying 'Y' below plot. Exact format of input can be found in the README files provided with 
each example included with the package.

\section{Parallelization and optimization}
\begin{figure}[!ht]
	 \centering
	   \includegraphics[scale=0.15]{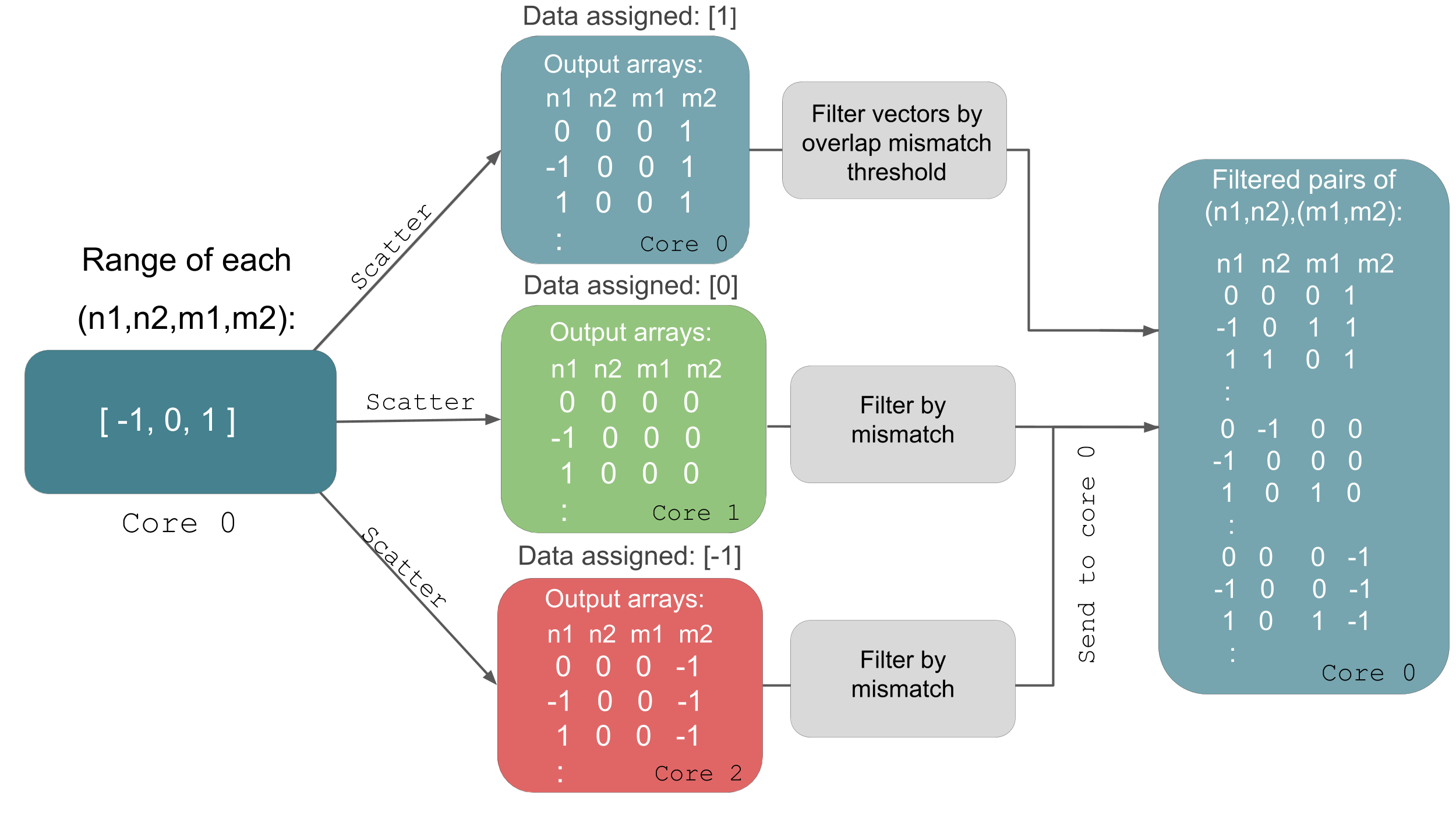}
	     \caption{ 
	     Schematic of the parallelization used in Twister to find commensurate twist angles.
	       }
	       \label{figure:6}
\end{figure}

The primary function for finding superlattice vectors in get\_ang.py is 
the funtion \emph{clt()}
and scales as O(N$^4$). We therefore optimise and parallelize 
this part of the code.
Consider the unit-cell lattice vectors of bottom layer                                                          
to be $\mathbf{a}_1$, $\mathbf{a}_2$, $\mathbf{a}_3$                                                            
and of top layer to be $\mathbf{b}_1, \mathbf{b}_2, \mathbf{b}_3$. Let superlattice in                          
bottom layer be given as $\mathbf{v}_1 = n_1\mathbf{a}_1 + n_2\mathbf{a}_2$, and top                            
layer as  $\mathbf{v}_2 = m_1\mathbf{b}_1 + m_2\mathbf{b}_2$. The function clt()                                
utilises user defined unit-cell lattice vectors and range of $n_1$,$n_2$,$m_1$,$m_2$ to checks 
all possible combinations of superlattice vectors for solutions. To parallelize computing 
the permutations, the range is scattered across the cores via MPI, using package 
\emph{mpi4py}. Each core gets $N_{nm}/N_c$ numbers from the range, 
where $N_{nm}$ is the length of range of $n_1,n_2,m_1,m_2$ and $N_{nm}$ is the user-defined number 
of cores. For each element in its assigned data, the core generates a four-columned \emph{NumPy} array, 
each column corresponding to $n_1,n_2,m_1,m_2$. The column for $m_2$ is uni-valued and 
only contains corresponding element in the data assigned to the core. The rows for $n_1,n_2,m_1$ 
contain all possible ordered triplets from the range of $n_1,n_2,m_1,m_2$ (with repeats 
allowed). These triplets were made using the \emph{product()} iterator from the \emph{Itertools} module 
of Python. A schematic for the the parallelized permutation calculations can be found in Figure \ref{figure:6}.

Using NumPy operations, the columns for $n_1,n_2$ are vector multiplied with $\mathbf{a}_1, \mathbf{a}_2$ 
and columns for $m_1, m_2$ are vector multiplied with rotated top layer vectors, 
$\mathbf{b}_1^r, \mathbf{b}_2^r$. The resulting NumPy arrays, $\mathbf{v}_1$ and $\mathbf{v}_2^r$, 
are used to calculate mismatch between them and the arrays are filtered according 
to the user-defined threshold (Eqn. \ref{eqn_clt_delta}). 
NumPy arrays for $\mathbf{v}_1, \mathbf{v}_2^r$, ($n_1, n_2$), ($m_1, m_2$) and mismatch between the 
pairs, are sent by different cores and received by the core with $rank=0$. All possible pairs of 
superlattice vectors are then made using \emph{combinations()} iterator from the \emph{Itertools} 
module is used to make all possible pairs of superlattice vectors. They are filtered for 
non-parallel pairs. No parallelization is 
applied in the remaining steps. NumPy operations and arrays are used for further filtering 
depending on angle between superlattice vectors, strain in lattice parameters, and area of supercell.

\section{Conclusion}

We present the Twister package to construct commensurate moir\'e superlattice for
any combination of 2D material. The package also streamlines  
atomic relaxation of the moir\'e superlattices by interfacing with classical forcefield
calculations implemented in the LAMMPS package. 
We demonstrate the construction of the superlattices for five test 
systems, twisted bilayer MoS$_2$, MoS$_2$/MoSe$_2$, 
graphene/hBN, hBN/MoSe$_2$ and twisted bilayer phosphorene. 
We also use the package to find commensurate twist-angles for the 
phosphorene bilayer and MoS$_2$/MoSe$_2$  
heterobilayer. The package is written in
Python and uses MPI parallelization for the computationally demanding steps.

\section{Acknowledgments}
The authors thank the Supercomputer Education and Research Centre 
(SERC) at IISc for providing the computational facilities.







\section*{References}
\bibliographystyle{elsarticle-num}







\end{document}